\documentclass[a4paper,aps,prd,showpacs,nofootinbib,preprintnumbers,amsmath,amssymb,mcite,superscriptaddress,12pt]{revtex4-1}

\usepackage[table]{xcolor}
\usepackage{graphicx}
\usepackage{dcolumn}
\usepackage{bm}
\usepackage{esvect}
\usepackage{accents}

\usepackage{hyperref}
\hypersetup{colorlinks=true,linkcolor=magenta,anchorcolor=green,citecolor=cyan,filecolor=black,menucolor=black,urlcolor=brown}
\usepackage{slashed}

\newcommand{\be}{\begin{equation}}
\newcommand{\ee}{\end{equation}}
\newcommand{\ba}{\begin{eqnarray}}
\newcommand{\ea}{\end{eqnarray}}

\begin{document}

\title{The Swampland and Screened Modified Gravity}

\author{Philippe Brax}
\affiliation{Institut de Physique Th\'eorique, Universit\'e  Paris-Saclay, CEA, CNRS, F-91191 Gif-sur-Yvette Cedex, France}
\author{Carsten van de Bruck}
\affiliation{School of Mathematics and Statistics, University of Sheffield, Hounsfield Road, Sheffield S3 7RH, United Kingdom}
\author{Anne-Christine Davis}
\affiliation{Department of Applied Mathematics and Theoretical Physics, Centre for Mathematical Sciences, Wilberforce Road,
Cambridge, CB3 0WA, United Kingdom}

\begin{abstract}
We consider the implications of the swampland conjectures on scalar-tensor theories defined in the Einstein frame in which the scalar interaction is screened. We show that chameleon models are not in the swampland provided the coupling to matter is larger than unity and the mass of the scalar field is much larger than the Hubble rate. We apply these conditions to the inverse power law chameleon and the symmetron. We then focus on the dilaton of string theory in the strong coupling limit, as defined in the string frame. We show that  solar system tests of gravity imply that viable dilaton models are not in the swampland. In the future of the Universe, if the low energy description with a single scalar is still valid and the coupling to matter remains finite, we find that the scalar field energy density must vanish for models with the chameleon and symmetron mechanisms. Hence in these models dark energy is only a transient phenomenon. This is not the case for the strongly coupled dilaton, which keeps evolving slowly, leading to a quasi de Sitter space-time.

\end{abstract}

\date{\today}

\maketitle


\section{Introduction}
The standard model of cosmology, the $\Lambda$CDM model, is an excellent description of current cosmological and astrophysical data. It requires two ingredients, which call for  physics beyond the standard model of particle physics: dark matter and dark energy. Dark matter is believed to be a particle appearing in theories beyond the standard model (BSM), while dark energy has yet to find a satisfactory explanation. The cosmological constant is the simplest candidate for dark energy and the data are consistent with it. It predicts that in the far future the universe will approach de Sitter space-time with a constant expansion rate. Theories which combine the principles of particle physics with that of General Relativity have yet to find an explanation for the origin of the cosmological constant such as a residual vacuum energy density. Recently it has been argued that de Sitter space-time cannot be realised in string theory, see e.g. \cite{Ooguri:2006in,Obied:2018sgi,Ooguri:2018wrx}, see \cite{Palti:2019pca} for a review and see \cite{Cicoli:2018kdo} for a word of caution about the swampland programme. If these results holds, then either string theory, as currently understood, is wrong or the current accelerated expansion is not due to a cosmological constant. Instead it would have to be driven by other degrees of freedom in the theory. The de Sitter and distance  conjectures, which we will summarise in the next section, put constraints on the effective low-energy theory of string theory. In particular the de Sitter conjecture strongly restricts  the slope of the potentials for such  scalar fields, which has huge implications for inflation and dark energy physics, for an incomplete list see \cite{Agrawal:2018own,Heisenberg:2018yae,Heisenberg:2018rdu,Marsh:2018kub,Akrami:2018ylq,Colgain:2018wgk,DAmico:2018mnx,Olguin-Tejo:2018pfq,vandeBruck:2019vzd,Agrawal:2019dlm,Hardy:2019apu,Baldes:2019tkl,Colgain:2019joh,Achucarro:2018vey,Kehagias:2018uem,Brahma:2018hrd,Garg:2018reu,Dimopoulos:2018upl,Kinney:2018kew,Kinney:2018nny,Aragam:2019khr}.
At low energy it is generally expected that, in the absence of underlying symmetry,  the scalar field responsible for the cosmic acceleration should be coupled to matter. For models of dark energy this follows from the quantum loops mediated by gravitons which couple dark energy and matter. In string theory, this is for instance the case of the string dilaton which couples universally to matter. Such universal couplings would naturally lead to violations of the solar system tests of gravity due to the presence of a fifth force modifying gravity significantly, hence ruling out most of these models as low energy candidates for a description of our Universe.

More generally, scalar fields which appear in string theory could be  coupled to different matter species with different strengths. As such the  couplings to dark matter are less constrained than the ones to standard model particles, simply because local tests of gravitation are not sensitive to dark matter per se. However, if these couplings to dark matter are constant,  cosmology bounds them in a stringent way \cite{vandeBruck:2017idm,Miranda:2017rdk}.  In the case of the interactions to the standard model particles, the coupling of the scalar fields is strongly constrained  by the Cassini experiment \cite{Bertotti:2003rm} when the force is long-ranged.  Such small couplings are not  natural unless they result from the dynamics of the models, i.e. if they follow from a screening mechanism \cite{Khoury:2010xi}. Coming back to the scalar field emanating from string theory whose evolution would generate the late time acceleration of the expansion of the Universe,  it seems highly relevant to investigate whether screening mechanisms, which would lead to a dynamical suppression of fifth force effects,  could be realized in the string theory context. In this paper, we will discuss three of such mechanisms studied so far in cosmology, namely the chameleon and symmetron mechanisms and the strongly coupled dilaton. These models are phenomenological, but they serve as a good playground for other screening mechanisms.
The strongly coupled dilaton is inspired from stringy considerations, i.e. the self-interaction potential of the runaway dilaton \cite{Gasperini:2001pc} and the least coupling principle
\cite{Damour:1994zq}. Let us briefly summarize the basics of the mechanisms here:

\begin{itemize}

\item In the chameleon mechanism, the mass of the scalar fields depends strongly on the environment \cite{Khoury:2003rn}. This is achieved by an interplay of the interactions with ambient matter and the self-interactions of the field. Examples of these theories include $f(R)$--theories which are consistent with local experiments.
\item In the symmetron mechanism, the potential of a scalar field is symmetry--breaking, whereas the conformal coupling is $Z_2$--invariant \cite{Hinterbichler:2011ca}. The coupling of the scalar field is field dependent. In regions of high density, the symmetry $\phi \rightarrow -\phi$ is unbroken, but in low density region, this symmetry is spontaneously broken. In dense environments the coupling to matter would vanish.
\item In the case of the strongly coupled dilaton, the potential of the scalar field is of exponential form, $V\propto e^{-\lambda\phi}$, in the string frame. The conformal coupling of the scalar field to matter possesses a minimum. In the absence of  the potential the field would be driven towards the minimum of the coupling function during the radiation and matter dominated areas, where the coupling of the scalar to matter would vanish (this mechanism has been called the "least-coupling principle" \cite{Damour:1994zq}.) The potential can be arranged such that the scalar field acts as a dark energy component \cite{Brax:2010gi}.	
\end{itemize}

All three mechanisms will be discussed in more detail below, with the emphasis on how these screening mechanism are compatible with the de Sitter and distance conjectures. As we will see, the swampland conjectures will put constraints on each of the individual screening mechanisms and hence on their possible realisations  in string theory. We will also discuss the validity of the theories as a description of the Universe in the far future.

The paper is organised as follows: In Section \ref{sec:conj} we recall the conjectures related to the swampland of string theory.  We also summarise some generic facts about scalar--tensor cosmology. In section \ref{sec:chameleon}, we  find a generic bound on the coupling between the scalar field responsible for dark energy and matter. In \ref{sec:chameleon},
\ref{sec:symmetron} and \ref{sec:dilaton} we study the implications of the de Sitter and distance conjectures on the chameleon, the symmetron and dilaton screening mechanisms, respectively. We summarise our findings and conclude in section \ref{sec:conclusions}.

\section{The de Sitter and distance conjectures in scalar-tensor cosmology}
\label{sec:conj}

\subsection{The conjectures}
De Sitter vacua are particularly hard to find in string theory. It has been recently conjectured that the vacuum of string theory is better described by
the dynamics of a scalar field whose potential must satisfy the inequality
\be
\left \vert \frac{\partial V}{\partial \phi} \right \vert \ge c \frac{V(\phi)}{m_{\rm Pl}},
\label{des}.
\ee
or the corresponding constraints on its mass
\be
\frac{\partial^2 V}{\partial \phi^2} \leq -c' \frac{V}{m_{\rm Pl}^2}~.
\ee
Here, $c$ and $c'$ are constants of order one. The distance conjecture states that the scalar field should not roll too far in field space, otherwise  low energy excitations would become relevant hence jeopardising the effective description of the vacuum being simply endowed with a single scalar field
\be
\Delta \phi \le d~m_{\rm Pl}
\label{dis}
\ee
where $d= {\cal O}(1)$ and $\Delta \phi$ is the total excursion of the scalar field between the very early Universe and now. These are constraints on the low--energy effective field theory allowed from string theory. If true, they imply that the current accelerated expansion of the universe is not due to a non-vanishing cosmological constant but driven by at least one degree of freedom in string theory.

\subsection{Scalar-tensor cosmology}

We are interested in consequences of the de Sitter and distance conjectures in scalar-tensor theories with a screening mechanism. Below we will recall a few useful facts on scalar-tensor theories which apply to all theories considered in this paper.

Scalar-tensor theories can be written in either the Einstein or Jordan frame. The Jordan frame metric is related to the Einstein frame metric by a conformal transformation of the form
\be
g_{\mu\nu}^J= A^2 (\phi) g_{\mu\nu}^E,
\ee
or equivalently, matter particles have a field dependent mass
\be
m= A(\phi) m_0
\ee
in the Einstein frame. The function $A(\phi)$ will differ for the different screening mechanism discussed in this paper.

The dynamics of the scalar field are governed by the effective potential
\be
V_{\rm eff}(\phi) = V(\phi) + (A(\phi)-1) \rho.
\ee
in the presence of non-relativistic matter of conserved energy density $\rho$. The Friedmann equation can be written as
\be
H^2 = \frac{ \rho_{\rm eff} +\rho}{3m_{\rm Pl}^2}.
\ee
The energy density $\rho_{\rm eff}= \frac{\dot \phi^2}{2}+V_{\rm eff}(\phi)$ plays the role of dark energy.
The conservation equation
\be
\dot \rho_{\rm eff} +3H (\rho_{\rm eff} +p_\phi)=0.
\label{conn}
\ee
where $p_\phi= \frac{\dot \phi^2}{2}-V_{}(\phi)$
implies that the dark energy equation of state is
\be
\omega_\phi= \frac{p_\phi}{\rho_{\rm eff}}.
\ee
Moreover the conservation equation (\ref{conn}) implies the Klein-Gordon equation
\be
\ddot \phi +3H \dot \phi +\frac{\partial V_{\rm eff}}{\partial \phi}=0
\ee
which depends on the effective potential. Notice that the effective potential depends on the conserved matter.

In the following we will take the potential $V(\phi)$ and the coupling functions $A(\phi)$ as the low energy results of dimensionally reducing extra dimensions, integrating out heavy fields and taking into account early Universe, i.e. high energy, phase transitions. As the de Sitter and distance conjectures are statements about the scalar fields in the low--energy field theory, we apply them to the potential $V(\phi)$ as this controls the existence of de Sitter space in empty space-time, i.e. when all matter in the Universe has been diluted by the cosmological expansion.

\section{The chameleon}
\label{sec:chameleon}

In the chameleon models the effective potential has a minimum $\phi(\rho)$ and the field tracks the minimum cosmologically \cite{Brax:2004qh}.
The condition for the minimum of the effective potential is
\be
\frac{\partial V}{\partial \phi}= -\beta A \frac{\rho}{m_{\rm Pl}}
\ee
where the coupling to matter is
\be
\beta\equiv m_{\rm Pl} \frac{\partial \ln A}{\partial \phi}.
\ee
We assume  without loss of generality that $\beta$ is positive. The field tracks the minimum provided the mass \cite{Brax:2004qh,Brax:2012gr}
\be
m^2= \frac{\partial^2 V_{\rm eff}}{\partial \phi^2}\vert_{\phi(\rho)}
\ee
is greater than the Hubble rate
\be
m\gg H.
\ee

\subsection{The original chameleon model}

Let us first look at the original chameleon model \cite{Brax:2004qh} before we move to a more general case which includes $f(R)$ gravity. In the original cosmological model for   chameleons, the potential is of the form
\be
V = \Lambda^4 e^{\left( \frac{\Lambda}{\phi} \right)^n} ~,
\ee
where $\Lambda$ is an energy scale of the order of the current dark energy scale. The function $A(\phi)$ is assumed to be of exponential form, i.e. $A(\phi) = \exp(\beta\phi/m_{\rm Pl})$ with $\beta\geq 0$ constant. Note that this model, at face value, does not comply with the de Sitter criterion, as $V\rightarrow \Lambda^4$ for $\phi \rightarrow \infty$. But according to the distance criterion we expect that this low--energy theory breaks down anyway for large field values, so we have to keep in mind that the chameleon model, if realised from fundamental physics, will become invalid at some point in the distant future. However, we will now show that the de Sitter conjecture puts a constraint on the coupling $\beta$. The field value at the minimum can be obtained as

\be
\phi^{n+1}= \left ( \frac{n V m_{\rm Pl}}{\beta \rho} \right )
\ee
Using this equation and  the de Sitter conjecture, we obtain a bound on the matter coupling:
\be\label{boundoriginal}
\beta \geq c \frac{V}{\rho},
\ee
or, using the cosmological density parameter, $\beta \geq c ~\Omega_{\rm DE} / \Omega_{\rm M}$. Hence, the coupling cannot be arbitrarily small in this model.

The discussion so far has assumed that the field sits in the minimum of the effective potential. If this is not the case, e.g.   in the very early radiation dominated epoch, the de Sitter conjecture implies
\be
\phi^{n+1} \leq \frac{c}{n} m_{\rm Pl} \Lambda^n ~.
\ee
This equation bounds the field value at any given time. In addition, the field is subject to kicks at times when species become non--relativistic, because at that point, the trace of the energy momentum tensor of the species no longer vanishes and contributes to the effective potential. Summing up all contributions it was found in \cite{Brax:2004qh} that the kicks can displace the field be an amount of order $\beta m_{\rm Pl}$. The distance conjecture would then imply that $\beta \leq d$. Together with eq. (\ref{boundoriginal}), this implies that $\beta$ has to be of the same order as the numbers $c$ and $d$ in the de Sitter and distance conjectures. To avoid a violent displacement of the chameleon, the field needs to settle at the minimum of the effective potential either during or shortly after inflation. We will come back to the issue of the initial conditions for the scalar field in the discussion at the end of the paper.

\subsection{A generic bound on the coupling}

We can derive a generic bound on the coupling which is applicable to more general scalar-tensor theories in the Einstein frame. To obtain this bound, it is convenient to write
\be
\frac{\dot \phi^2}{2}= V(\phi) + \omega _\phi \rho_{\rm eff}
\ee
such that
\be
V(\phi)= \rho_{\rm eff} -\frac{\dot \phi^2}{2}- (A-1) \rho= (1-\omega_\phi) \rho_{\rm eff} - V(\phi) - (A-1) \rho
\ee
from which we have
\be
V(\phi) = (1-\omega_\phi) \frac{\rho_{\rm eff}}{2} - (A-1) \frac{\rho}{2}
\ee
Using the minimum equation and the distance conjecture (\ref{dis}) we find that the de Sitter constraint (\ref{des}) gives
\be
\beta \ge \frac{c(1-\omega_\phi)}{2} \frac{\rho_{\rm eff}}{\rho}~,
\ee
where we have normalised $A(\phi)$ to be close to unity now and expanded $A(\phi)= 1+ \beta \frac{\phi}{m_{\rm Pl}}$ as the field $\phi$ must have a short excursion in field space.
Now we are interested in models where $\rho_{\rm eff}$ represents the dark energy component of the Universe. We assume that it grows monotonically whereas $\rho$ decreases in the cosmic history (as it is the case for the original chameleon model as well as for $f(R)$ theories). Hence the most stringent constraint is
\be
\beta \ge \frac{c(1-\omega_\phi)}{2} \frac{\Omega_{\Lambda0}}{\Omega_{m0}}~,
\label{bound1}
\ee
where $\Omega_{\Lambda 0}$ and $\Omega_{m0}$ are the dark energy and matter proportions now, i.e. $\frac{\Omega_{\Lambda0}}{\Omega_{m0}}\simeq 3$. This generalises (\ref{boundoriginal}). For $\omega_\phi \simeq -1$
this implies that the coupling to matter cannot be small in general chameleon models. This puts pressure on models of the $f(R)$ type, in which $\beta=\frac{1}{\sqrt 6}$ as soon as $c\gtrsim 1$.

\subsection{Screening and the distance conjecture}

Assuming that the field at the minimum vanishes in dense environments and parameterising $\rho= \rho_0/a^3$ as a function of $a$ as in cosmology, the field at the minimum can be evaluated exactly using \cite{Brax:2012gr}
\be
\frac{\phi(\rho)}{m_{\rm Pl}}= 3 \int_0^a \frac{dx}{x} \frac{\rho (x) \beta (x)}{m_{\rm Pl}^2 m^2(x)}
\ee
provided the dependence of $m(a)$, i.e. the mass of the field at the minimum,  and $\beta(a)$, i.e. the coupling to matter at the minimum, are known. This corresponds to the full excursion $\Delta \phi (a)$ in this models as we have assumed here that when $\rho$ becomes infinite the field at the minimum converges to zero.
Writing $m= m_0 \tilde m(a), \ H(a)= H_0 h(a)$ where $h(a)$ and $\tilde m (a)$ are two functions of $a$ of order one we obtain that
\be
\frac{m^2_0}{H_0^2}\ge \frac{9\Omega_{m0}}{d} I(a)
\ee
where
\be
I(a)= \int_0^a \frac{dx}{x} \frac{h^2 (x) \beta (x)}{\tilde m^2(x)}
\ee
is a function of order one, see \cite{Brax:2011aw}. Now the tracking of the minimum requires that $m^2_0/H_0^2 \gg 1$, implying that the distance conjecture is always satisfied now.
In the future when $a\to \infty$, and assuming that $m(a)\gg H(a)$ to guarantee the tracking behaviour, if the integral $I(a)$ remains bounded then the distance conjecture remains valid for all times.

\subsection{Solar systems tests of gravity and the swampland}

Before we conclude this Section, we will briefly discuss constraints from solar system gravity tests and the implications for the swampland conjectures. We refer to the appendix for more details.

The Cassini and Laser Lunar Ranging tests of respectively fifth forces and the strong equivalence in the solar system imply bounds on the excursion of the scalar field in galactic environments similar to the Milky Way
\be
\Delta \phi_G= \phi_G- \phi_c \leq 10^{-15} m_{\rm Pl}
\ee
which is well within the Planck scale. Here $\phi_c$ is the value of the field in dense matter, which differs from zero for the dilaton. Similarly the Cassini bound on the existence of fifth forces for nearly massless scalar fields imply that
\be
\beta_G \lesssim 10^4,
\ee
hence the coupling to  matter in the Milky Way cannot be exceedingly large.
Together with the bound (\ref{bound1}), this implies that the coupling to matter is both bounded from below and from above.

\section{The symmetron}
\label{sec:symmetron}

The cosmological symmetron is a model where a scalar field undergoes a $Z_2$ breaking  transition at low energy. In the symmetric phase, the coupling of the scalar field to matter vanishes whilst it is non-vanishing in the symmetry-breaking phase.  The potential for these models is Higgs-like with
\be
V(\phi)= V_0 -\frac{\mu^2}{2} \phi^2 + \frac{\lambda}{4} \phi^4.
\ee
The value of $V_0$ has to be chosen to lead to the acceleration of the expansion of the Universe.
The coupling function determining the coupling  to matter differs from the one of the original inverse power law chameleon and is simply a quadratic function around the origin
\be
A(\phi)= 1 +\frac{\phi^2}{2M^2}.
\ee
This has to be seen as an expansion in powers of $\phi/M$.
The de Sitter conjecture implies that $\mu^2 m_{\rm Pl}^2 > c' V_0$ and therefore  $\mu\gtrsim H_0$. This is a very weak condition.

The coupling to matter is
\be \label{symmetroncoupling}
\beta (\phi)=\frac{m_{\rm Pl}}{M^2} \phi
\ee
which is linear in the field as long as $\phi \ll M$.
When $\rho>\mu^2 M^2$, the minimum of the effective potential is at the origin and the coupling to matter vanishes. Otherwise the minimum is at
\be
\phi(\rho)= \frac{\sqrt{ \mu^2 -\frac{\rho}{M^2}}}{\sqrt \lambda}.
\ee
We require that $\mu \ll \sqrt \lambda M$ which guarantees that $\phi\ll M$.
The cosmological symmetron is such that the $Z_2$ breaking occurs in the recent past implying that
\be
\mu M \simeq H_0 m_{\rm Pl}.
\ee
The vacuum value of the coupling to matter is given by
\be
\beta_0= \frac{\mu m_{\rm Pl}}{\sqrt \lambda M^2}.
\ee
The de Sitter conjecture implies that
\be
\frac{\rho}{M^2} \phi(\rho) \ge c \frac{V(\phi)}{m_{\rm Pl}}.
\ee
Assuming that the symmetron leads to dark energy now, we get
\be
\beta_0 \ge c \frac{\Omega_{\Lambda 0}}{\Omega_{m0}}
\ee
which is another instance of the generic bound (\ref{bound1}) when the equation of state of dark energy is close to $-1$.

At high density, the field is at the origin due to the coupling to matter. In the present Universe we find that
\be
\frac{\Delta \phi}{m_{\rm Pl}} \le \frac{\mu}{\sqrt \lambda m_{\rm Pl}} \ll  \frac{\mu}{\sqrt \lambda M}\lesssim 1
\ee
as long as $M\ll m_{\rm Pl}$ and the last step comes from requiring that $\phi \ll M$ for the validity of the $\phi$ expansion in $A(\phi)$.
As outlined in the previous section and in more detail in the appendix, tests of the equivalence principle imply that in the Milky-Way $\phi_G \le 10^{-15} m_{\rm Pl}$. When  $\rho_G\simeq 10^6 \rho_{m0}$ is assumed to be larger than $\mu^2 M^2$, we have $\phi_G=0$ and the distance conjecture is satisfied. When the symmetry breaking happens at larger density, then $\phi_G \le \mu/\sqrt \lambda $ and we must require that $\mu \le 10^{-15} \sqrt \lambda m_{\rm Pl}$. As long as $\lambda$ is not tiny, the interval $H_0 \lesssim \mu \le 10^{-15} \sqrt \lambda m_{\rm Pl}$ is not empty. The de Sitter conjecture implies that
\be
M^2 \le  \frac{\Omega_{m0}}{c\Omega_{\Lambda 0}} \frac{\mu m_{\rm Pl}}{\sqrt \lambda} \lesssim 10^{-15} m_{\rm Pl}^2
\ee
which guarantees that $M\ll m_{\rm Pl}$. Hence the symmetron is not in the swampland as long as the coupling to matter in the present Universe is large enough.

\section{The strongly coupled dilaton}
\label{sec:dilaton}

So far we have dealt with scalar-tensor theories where the potential $V(\phi)$ is defined in the Einstein frame.
In this section we are interested in a string-inspired model where the scalar field $\phi$ corresponds to the dilaton associated with the string coupling constant \cite{Brax:2010gi}. The model is naturally defined in the string frame. We briefly review the model in the following.

In the string frame the dilaton action reads
\be
\begin{aligned} \mathcal{S}=& \int \sqrt{-\tilde{g}} d^{4} x\left[\frac{e^{-2 \psi(\tilde \phi)}}{2 l_{s}^{2}} \tilde{R}+\frac{Z(\tilde \phi)}{2 l_{s}^{2}}\left(\tilde{\nabla}^{2} \tilde \phi\right)-\tilde{V}(\tilde\phi)\right] \\ &+\mathcal{S}_{\mathrm{m}}\left(\Psi_{i}, \tilde{g}_{\mu \nu} ; g_{i}(\tilde\phi)\right) \end{aligned}
\ee
where $l_s$ is the string length, $\Psi_i$ are the matter fields and the $g_i$ are coupling constants which depend on the dilaton $\tilde \phi$. We can bring this action into the Einstein frame in which we have performed our analysis so far. We define the Einstein metric $g_{\mu\nu}$ by
\be
\tilde{g}_{\mu \nu}=A^{2}(\tilde \phi) g_{\mu \nu},
\ee
where the coupling function is given by
\be
A(\phi)=l_{\mathrm{s}} e^{\psi(\tilde \phi)} / \kappa_{4}
\ee
and the gravitational coupling is given by
$
\kappa_{4}^{2}=8 \pi G_{N}.
$
We have the freedom to normalise $A(\tilde \phi)$ such that $A(\tilde \phi_0)=1$ now where $\tilde \phi_0$ is the value of the dilaton cosmologically now. We introduce
$
c_{1} \equiv l_{\mathrm{s}} / \kappa_{4}=\exp \left(-\psi\left(\phi_{0}\right)\right).
$
The kinetic terms are now dependent on
\be
k^{2}(\tilde \phi)=3 \beta^{2}(\tilde  \phi)-A^{2}(\tilde \phi) Z(\tilde \phi) / 2 c_{1}^{2}
\ee
where
\be
\tilde \beta(\phi) = (\ln A)_{,\tilde \phi}
\ee
is the coupling to matter for the unnormalised field $\tilde \phi$.
The resulting action becomes
\be
\begin{aligned} \mathcal{S}=& \int \sqrt{-g} d^{4} x\left(\frac{R(g)}{2 \kappa_{4}^{2}}-\frac{k^{2}(\tilde \phi)}{\kappa_{4}^{2}}(\nabla \tilde \phi)^{2}-V(\tilde \phi)\right) \\ &+\mathcal{S}_{\mathrm{m}}\left(\Psi_{i}, A^{2}(\tilde \phi) g_{\mu \nu} ; \tilde \phi\right) \end{aligned},
\ee
where the potential is
\be
V(\tilde \phi)=A^{4}(\tilde \phi) \tilde{V}(\tilde \phi).
\ee
In the strong coupling limit when $\tilde \phi$ is large we will assume that
\be
\begin{aligned} \tilde{V}(\tilde \phi) & \sim {V}_{0} e^{-\tilde \phi}+\mathcal{O}\left(e^{-2 \tilde \phi}\right) \\ Z(\tilde \phi) & \sim-\frac{2 c_{1}^{2}}{\lambda^{2}}+b_{Z} e^{-\tilde \phi}+\mathcal{O}\left(e^{-2 \tilde \phi}\right) \\ g_{i}^{-2} & \sim \overline{g}_{i}^{-2}+b_{i} e^{-\tilde \phi}+\mathcal{O}\left(e^{-2 \phi}\right) \end{aligned}
\ee
The constants are assumed to be such that $b_Z\simeq b_i ={\cal O}(1)$ and $c_1$ can vary between unity and ${\cal O}(1 / l_s m_{\rm Pl})$, where the string scale is generically taken to be $l_s\gg l_{\rm Pl}=1/m_{\rm Pl}$. In the strong coupling regime we expect thus
\be
k(\tilde \phi) \approx \lambda^{-1} \sqrt{1+3 \lambda^{2} \beta^{2}(\tilde \phi)}.
\ee
which depends on the coupling to matter. It is useful to normalise the field  to connect with the other sections of this paper. We now define
\be
\kappa_4 d\phi=\sqrt 2 k(\tilde \phi) d \tilde \phi
\ee
The effective potential which governs the evolution of $\phi$ is given by
\be
V_{\mathrm{eff}}\left(\phi \right)=V_{0} A^{4}(\tilde\phi) e^{-\tilde \phi}+(A(\tilde\phi)-1)\rho
\ee
in the presence of non-relativistic matter.
Notice the crucial factor of $A^4$ in the matter-less part of the potential.
The minimum of the potential is obtained for
\be
\tilde \beta\left(\tilde \phi_{\min }\right)=\frac{V\left(\tilde \phi_{\min }\right)}{A\left(\tilde \phi_{\min }\right) \rho_{\mathrm{m}}+4 V\left(\tilde \phi_{\min }\right)}
\ee
which is an equation for $\tilde \phi_{\min}$. Notice that $\tilde \beta\left(\tilde \phi_{\min }\right)\le \frac{1}{4}$.
Moreover, the fact that the theory is originally defined in the string frame will modify the bound on the coupling to matter that we will find below.

The coupling to gravity $\beta$ of the normalised scalar field is defined by
\be
\beta (\phi)= \frac{\tilde \beta (\tilde \phi)}{\sqrt 2 k(\tilde \phi)}.
\ee
We have the relation
\be
2 \beta^2= \frac{\tilde \beta^2}{3 \tilde \beta^2 + \frac{1}{\lambda^2}}.
\ee
Gravitational tests  in the solar system require that $\beta \ll 1$ which cannot be achieved if $\tilde \beta \lambda \gtrsim 1$, as then $\beta \simeq \frac{1}{\sqrt 6}$. Tests of gravity can only be passed when $\lambda \tilde \beta \ll 1$, i.e. $\lambda$ is bounded from above. In this case we have
\be
\phi \simeq \frac{\sqrt 2 }{\lambda} m_{\rm Pl} \tilde \phi
\ee
and the potential becomes
\be
\tilde V \simeq V_0 e^{- \lambda \phi/\sqrt 2 m_{\rm Pl}}.
\ee
Similarly the coupling to gravity is then
\be
\beta \simeq \frac{\lambda}{\sqrt 2} \tilde \beta.
\ee
We will assume that the least coupling principle applies in the recent past of the Universe and expand $A(\tilde\phi)$ around its minimum taken to be the value of field in very dense environments $\tilde\phi_0$.
\be
A(\tilde \phi) = 1 + \frac{A_2}{2} (\tilde \phi -\tilde \phi_0)^2 + \dots
\ee
where the neglected terms are higher powers of $(\tilde \phi -\tilde \phi_0)$.
As the conformal factor $A$ deviates very little from unity in the late-time Universe, we can identify the dark energy scale with
\be
\rho_\Lambda\simeq V_0e^{-\phi_0}
\ee
 The minimum equation implies that in a dense environment we have
\be
A_2(\tilde \phi_{\min}-\tilde \phi_0)= \frac{V\left(\tilde \phi_{\min }\right)}{A\left(\tilde \phi_{\min }\right) \rho+4 V\left(\tilde \phi_{\min }\right)}.
\ee
In  dense environments such as the matter and radiation epochs the field value is essentially given by $\tilde \phi_0$, whilst at late time we have the approximation
\be
\tilde \phi_{\min} - \tilde \phi_0 \simeq \frac{1}{A_2} \frac{\rho_\Lambda}{\rho + 4 \rho_\Lambda}.
\label{mindil}
\ee
This is also related to the excursion $\Delta \phi$ of the field since the early Universe
\be
\frac{\Delta \phi}{m_{\rm Pl}} \simeq  \frac{\lambda}{\sqrt 2 A_2} \frac{\rho_\Lambda}{\rho + 4 \rho_\Lambda}\le \frac{\lambda}{4\sqrt 2 A_2}
\ee
as the value of the field in very dense matter is $\phi_0$.
Using that $A(\tilde \phi)$ is close to unity we also find the constraint from the de Sitter conjecture
\be
\lambda \ge \sqrt 2 c
\label{dedil}
\ee
which is compatible with the range of values of $\sqrt 2 c \le \lambda\le c_1$, i.e. when $l_s\gtrsim l_{\rm Pl}$.

As the value of the dilaton in very dense region does not vanish, the Lunar Ranging constraint reads
\be
\vert \phi_G -\phi_0\vert\simeq \frac{\lambda}{\sqrt 2 A_2} \frac{\rho_\Lambda}{\rho_G} m_{\rm Pl} \le 10^{-15} m_{\rm Pl}
\ee
as $\phi\simeq \phi_0$ inside matter. This implies that
\be
\frac{A_2}{\lambda}\ge 10^9
\ee
and the excursion (\ref{mindil}) is extremely small in Planck units as $A_2/\lambda$ is so large. Moreover in this regime the mass of the dilaton cosmologically is
\be
m_0 \simeq  \frac{\sqrt A_2}{\lambda} H_0
\ee
which is always large enough to guarantee that the dilaton tracks the minimum of the effective potential.
Coming back to the value of the coupling and using (\ref{dedil}) we find that
\be
\beta \gtrsim c \frac{\rho_\Lambda}{\rho+ 4\rho_\Lambda}
\label{newbound}
\ee
which is a weaker version than the generic bound we obtained previously (in eq. (\ref{bound1})). The main change comes from the $4\rho_\Lambda$ term in the denominator which comes from the fact that the dilaton potential is defined in the string frame and not in the Einstein frame. Thus, the strongly coupled dilaton does not violate the de Sitter and the distance conjectures when the string scale is significantly lower than the Planck scale. Note that dark energy is eternal as the scalar field approaches ${\tilde\phi}_{\rm min}$ but never reaches it (see eq. (\ref{mindil})).

\section{Discussion and Conclusion}
\label{sec:conclusions}

In this paper we have discussed the implications of the swampland on three screened modified gravity theories, namely the chameleon, the symmetron and the strongly coupled dilaton. In these theories, the dark energy scalar is universally coupled to matter, and hence producing a fifth force which needs to be hidden by a screening mechanism. While some of the screening mechanisms are meant to be only effective descriptions, which are  not valid for all values of the scalar field, our considerations have implications for each of the models. Let us summarise the findings for each of these theories separately:

\begin{itemize}
\item Since the chameleon field tracks the minimum of the effective potential for most of the cosmological history, the derivative of the potential is related to the matter density and the coupling between the chameleon field and matter. The distance and de Sitter conjectures then imply a {\it lower} bound on the coupling (eq. (\ref{bound1})). Note that this bound is time-dependent and strictly speaking we require it to be only valid up to the present epoch. The ratio $\rho_{\rm eff}/\rho$ will grow over time and larger values of $\beta$ are required. One expects that the field excursion over the cosmic history will eventually exceed one Planck unit at which point the theory will cease to be valid, even  probably before this time. Alternatively, the field will stop tracking the minimum of the effective potential in the very near future. Moreover the original chameleon model can only be an effective description of the universe up to the present epoch, as the potential energy does not vanish for arbitrary large field values and the universe approaches de Sitter space-time. If the field description does not break down in the future, the chameleon models must be modified with a vanishing potential asymptotically. Hence in these models, dark energy can only be transient.

\item Like the chameleon, the symmetron tracks the minimum of the effective potential for most of the cosmological history. The coupling of the symmetron is linear in the field (see eq. (\ref{symmetroncoupling})). The distance conjecture is easily fulfilled and again we find that the de Sitter conjecture implies that the coupling has to be large enough for the symmetron not to be in the swampland. In the future of the Universe, the symmetron will converge to a finite value well below the Planck scale. The bound on the coupling to matter (\ref{bound1}) implies that the minimum of the potential in vacuum must vanish, hence adjusting the constant $V_0$ in the potential. As in the chameleon model, in the symmetron model dark energy is only transient.

\item The strongly coupled dilaton, contrary to chameleons and symmetrons, is best defined in the string frame. The action in the Einstein frame is then derived, implying that the bound on the coupling to matter (\ref{newbound}) is modified compared to (\ref{bound1}) as obtained for all chameleon-like theories defined in the Einstein frame. When the least coupling principle is satisfied we find that the strongly coupled dilaton tracks the minimum of its effective potential. In field space, its excursion is always finite and of small magnitude in Planck units. As the field keeps evolving,  the cosmology of space-time is the one of a quasi de Sitter Universe. Contrary to chameleons and symmetrons, dark energy is eternal.

\end{itemize}

In a similar vein, we can discuss the initial conditions for the three types of models. Indeed we have assumed that the field sits at the minimum of its effective potential since early times. Once at the minimum, the condition on the mass of the scalar field $m\gg H$ guarantees that the field tracks the time-evolving minimum. In each of the three mechanisms let us discuss how the field could be attracted to the  minimum of the effective potential:

\begin{itemize}

\item For chameleon models \cite{Brax:2004qh} such as the inverse power law chameleon, the effective potential possesses a minimum during the inflationary era as the trace of the energy-momentum tensor of the inflaton is non-vanishing and nearly constant. The field falls exponentially fast towards the nearly-static minimum. When inflation stops and assuming that reheating is quasi-instantaneous, the minimum of the effective potential evolves rapidly towards a much larger value than during inflation. The field then starts evolving fast and overshoots the minimum before stopping after an excursion of around $\sqrt{6\Omega_\phi^i}m_{\rm Pl}$ where $\Omega_\phi^i$ is the initial energy fraction in the scalar, i.e. a small number. Notice that the field stops short of the Planck scale. Subsequently in the radiation era, the field is kicked by a negative fraction of the Planck scale every time a species decouples. This should eventually bring back the field within the basin of attraction of the minimum where it will eventually settle. The validity of this scenario has been questioned in \cite{Erickcek:2013dea}. In the absence of a concrete model of reheating, it is far more conservative to assume that the field sits at the minimum after reheating. This protects the field from being kicked during the decoupling of species.

\item For symmetrons, at high density, i.e. during inflation and after reheating, the field sits at the origin. When the matter density decreases, the field follows the minimum
\cite{Brax:2011pk}. In this model, there is no initial condition problem as the minimum is not shifted from its position during inflation, i.e. at the origin, to a new position in the early radiation era.

\item For dilatons the situation is similar to the one for symmetrons, i.e. very early in the Universe the field sits at the minimum of the coupling function. As the energy density of matter decreases, the field evolves with the minimum.

\end{itemize}

To conclude, we have shown that the de Sitter and distance conjectures have important implications for all three screening mechanism. In the case of chameleons, we find that $f(R)$ models come under pressure from the de Sitter conjecture, at least as long as the scalar field tracks the minimum of the effective potential (see \cite{Benetti:2019smr} on a different view of $f(R)$ gravity and the swampland). The lower bound on the coupling (\ref{bound1}) implies that those theories can not hold for arbitrarily long into the future. Like the original chameleon model, the theory will have to break down at some point (or the field no longer tracks the minimum of the effective potential). For example, other corrections to the Einstein--Hilbert action may become important. Moreover as shown in \cite{Hertzberg:2018suv} and elaborated in an appendix, the quantum corrections to the screened models do not lead to more fine-tuning than the usual cosmological constant problem provided one considers them as low energy effective theories below a cut-off scale of order $10^{-2}$ GeV. This is the low energy regime of cosmology where screening should take place, i.e. from Big Bang Nucleosynthesis onwards.

Given the implications of the swampland for dark energy physics, it seems highly relevant  to study the consequences of couplings of the scalar field to matter within string theory. This coupling  can be either  universal  to all forms of matter or only to one  sector, such as dark matter. Given the theoretical difficulties of constructing quintessential models within string theory \cite{Hertzberg:2018suv,Cicoli:2018kdo,Hebecker:2019csg}, the swampland conjectures   lead us to surmise that coupled models with screening mechanisms should play a role within string theory. The chameleon models with a constant coupling is difficult to construct within N=1 supergravity \cite{Brax:2006np} (see also \cite{Hinterbichler:2010wu} for an alternative point of view). They  are also under pressure from the de Sitter and distance conjectures. Furthermore, it has been argued that the form of the potential energy of the scalar field should be  related to the tower of particles via the Gibbons--Hawking (GH) entropy \cite{Ooguri:2018wrx}. If this is the case, then screening via the chameleon mechanism might not be possible. For example, if the mass of particles depends exponentially on the field then the GH entropy suggests that the potential energy of the scalar does as well; in such a setup the thin--shell mechanism in chameleon theories does not exist \cite{Brax:2010gi}. Alternatives such as field dependent couplings may be promising as hinted by the strongly coupled dilaton (there are also examples of chameleon theories with field dependent couplings, see e.g. \cite{Brax:2010kv}; these theories need to be studied in more detail). In particular, once solar system constraints on gravity are imposed, the strongly coupled dilaton keeps evolving without violating the distance conjecture and its potential energy leads to a quasi de Sitter space-time which evades Weinberg's no-go theorem \cite{Weinberg:1988cp}. A more thorough investigation of the strongly coupled dilaton from the string theory point of view would certainly add to this discussion.

\hspace{0.5cm}

\noindent {\bf Acknowledgements:} We are grateful to Yashar Akrami, Eran Palti and Cumrun Vafa for helpful comments regarding the de Sitter conjecture and to Mark Trodden for comments regarding quantum corrections. CvdB is supported in part by the Lancaster-Manchester-Sheffield Consortium for Fundamental Physics under STFC grant ST/P000800/1. ACD acknowledges partial support from STFC under grants ST/L000385 and
ST/L000636. This work is
supported in part by the EU Horizon 2020 research and innovation
programme under the Marie-Sklodowska grant No. 690575. This article is
based upon work related to the COST Action CA15117 (CANTATA) supported
by COST (European Cooperation in Science and Technology).  This work was made possible by Institut Pascal at University Paris-Saclay with the support of the P2I and SPU research departments and  the P2IO Laboratory of Excellence (program “Investissements d’avenir” ANR-11-IDEX-0003-01 Paris-Saclay and ANR-10-LABX-0038), as well as the IPhT.

\section{Appendix: Solar system gravity tests}
In this Appendix we briefly summarize constraints coming from solar system experiments.

\subsection{Strong equivalence principle}

The screening models lead to a violation of the strong equivalence principle for screened bodies. Contrary to point particles which couple to the scalar field with the coupling $\beta (\phi)$, extended bodies couple with a scalar charge
\be
\beta_{\rm eff}= \frac{\vert \phi_{\rm out}- \phi_{\rm in}\vert}{2 m_{\rm Pl } \Phi}
\ee
where $\Phi$ is the Newton potential at their surface.
These objects are screened when
\be
\beta_{\rm eff} \le \beta (\phi_{\rm out})
\ee
where $\phi_{\rm in}$ is the field value deep inside the body corresponding to the field value associated to the density of the object, and $\phi_{\rm out}$ is the field value far away from the object associated to the density of the environment. For most chameleon models, $\phi$ decreases with $\rho$ in such a way that we can approximate
\be
\beta_{\rm eff}= \frac{\vert \phi_{\rm out}\vert}{2 m_{\rm Pl } \Phi}
\label{eff}
\ee
which depends both on the environment and on the inner gravity of the object. For dilatons, $\phi_{\rm in}$ has to be kept in the previous expression.
Three screened bodies $A$, $B$ and $E$ embedded in the same background but with differing Newton potentials couple differently to the scalar implying a non-zero value for the
E\"otvos parameter
\be
\eta_{AB}= \frac{\vert a_A-a_B\vert}{\vert  a_A+a_B\vert}\simeq \beta_E \vert \beta_A-\beta_B\vert
\ee
where $a_{A,B}$ are the accelerations towards $E$. In the Moon-Earth-Sun system
 and the couplings depends on the objects as in (\ref{eff}), the constraint given by the Laser Lunar Ranging experiment on the violation of the equivalence principle for the Earth and the Moon falling towards the Sun is \cite{Khoury:2003rn}
\be
\beta_\oplus \le 10^{-6}.
\label{llr}
\ee
As $\Phi_\odot =10^{-9}$, this implies for the screened field in the Milky Way
\be
\phi_G \lesssim 10^{-15} m_{\rm Pl}.
\ee
Hence as long as the density dependence of $\phi(\rho)$ is not too strong and using $\rho_G\simeq 10^6 \rho_0$, we find that the distance conjecture is always satisfied for screened models which pass the LLR test.

\subsection{The Cassini experiment}

The Cassini satellite has given a strong constraint on long range forces in the solar system \cite{Bertotti:2003rm}. Assuming that the Compton wavelength  of the screened scalar in the solar is larger than the solar system, the deviation from Newton's law (or the Shapiro effect) implies that
\be
\beta_{\rm sat} \beta_\odot \le 10^{-5}.
\ee
Assuming that the satellite is not screened as it is a small object and using $\Phi_\odot =10^{-6}$ implying that $\beta_\odot \le 10^{-9}$ from (\ref{llr}), this leads to
\be
\beta_G \le 10^4.
\ee
Hence the constraint from the Cassini experiment on the coupling in the galactic environment is quite loose. It is certainly compatible with (\ref{bound1}) when the density dependence of $\beta$ is weak.

\section{Appendix: Quantum corrections}

We have focussed on  classical properties of scalar-tensor theories with screening. In this appendix, we will discuss the quantum corrections in these models. We will face the usual fine-tuning of the vacuum energy at low energy which requires one fine-tuning using a bare cosmological constant as a counter-term. Other quantum corrections are also important and will give  a restriction on the quantum validity of the models.

Let us concentrate on the matter contributions to the quantum corrections following \cite{Pietroni:2005pv}. In the Jordan frame, matter quantum corrections to the vacuum energy do not involve the scalar field at all and come from the vacuum diagrams with matter particles running in the loops. The result is formally divergent and equal to $\Lambda_{\rm qu}^4 (\mu) $ after regularisation and renormalisation. For instance in dimensional regularisation, the contributions involve quartic powers of the masses of particles up to logarithmic corrections which depend on the sliding scale $\mu$. In the Einstein frame, this would lead to a new potential $\delta V(\phi)= \Lambda_{\rm qu}^4(\mu) A^4(\phi)$. In general, $\Lambda_{\rm qu}(\mu)$ is much bigger than the dark energy scale. This is simply the usual cosmological constant problem. At the quantum level, one can always require that  the  bare cosmological constant $\Lambda_{\rm bare}$ whose role is to cancel  the infinities of the quantum corrections would also absorb  the finite part for a given value $\mu=\mu_0$. For this value of the sliding scale, the dark energy potential $V(\phi)$ is not corrected by quantum effects. This requires  the same fine-tuning as in all models of dark energy when facing the cosmological constant problem.

The quantum corrections to the potential $V(\phi)$ coming from the scalar itself have for magnitude $\delta V(\phi) \simeq {m^4_\phi}$ which is negligible as long as $m_\phi \ll 10^{-3}$ eV as required for dark energy scalar to have some influence on cosmological scales. Matter-scalar mixing can also lead to new constributions. For instance at two loops with one insertion of a scalar propagator, a fermion loop gives a contribution of order
\be
\delta V \simeq \beta^2 \frac{m^6_\psi}{m_{\rm Pl}^2}
\ee
which, for $\beta\gtrsim 1$, is a negligible correction to the late-time dark energy when $m_\psi \ll 10^{-2}$ GeV \cite{Hertzberg:2018suv}. As a result, screened models of dark energy are only low energy effective field theories with a low cut-off. Notice that this does not preclude the use of these models at low energy since Big Bang Nucleosynthesis which takes place around the energy scale of the order of the electron mass.

Finally we must analyse the quantum corrections to the coupling to matter $\beta$. When scalar and gravitational non-linearities are neglected it has been argued in \cite{Hui:2010dn} that the coupling $\beta$ receives only corrections from the wave-function renormalisation of the scalar $\phi$ by matter loops. The wave function renormalisation is  $Z_\phi \simeq 1+ \delta Z_\phi$ inducing a correction $\delta \beta\simeq  -\frac{1}{2} \beta \delta Z_\phi$ to $\beta$. At leading order for a fermion of mass $m_\psi$, we have
\be
\delta Z_\psi \simeq \beta^2 \frac{m^2_\psi}{m^2_{\rm Pl}}
\ee
which is negligible when $\beta\simeq 1$ at low energy. Mixing between the scalar and gravitons lead to a logarithmic correction to $\beta$ from a graviton loop
\be
\delta \beta \simeq \frac{m_\phi^2 \phi}{m^3_{\rm Pl}}\lesssim d\frac{m_\phi^2 }{m^2_{\rm Pl}}
\ee
using the distance conjecture. This is very small. Finally scalar loops give contributions in
\be
\delta \beta \simeq m_{\rm Pl} A'' V_{\rm eff}'''
\ee
which involves the triple derivative of the effective potential with respect to $\phi$ at the minimum of the effective potential.
This can be estimated using the tomographic map as
\be
 A'' \simeq \frac{A}{m^2_{\rm Pl}} ( \frac{d\ln \beta}{da} + \beta )\frac{m^2_\phi}{H^2}, \ \
V'''\simeq \frac{1}{\beta m_{\rm Pl}} \frac{m^2_\phi}{H^2} \frac{dm^2_\phi}{da}.
\ee
This leads to competing factors. Dimensionally we have $\frac{dm^2_\phi}{da} \sim m^2_\phi \ll m_{\rm Pl}^2$ which cannot be compensated by $\frac{m^4_\phi}{H^4}$ unless in extremely dense environments. As a result the correction to $\beta$ is negligible.

In conclusion, the quantum corrections are no worse than in usual quintessence models as long as the models are used at low energy below $10^{-2}$ GeV.

\bibliography{dmdebib}

\end{document}